\documentclass[aps,prl,twocolumn]{revtex4}

\usepackage{epsfig}
\usepackage{graphics}
\begin{document}
\title
{\bf Quantum Interference Effects in Electronic Transport 
through Nanotube Contacts}

\author{Calin \surname{Buia}\email[Email address]{: buia@physics.unc.edu}} 
\author{Alper \surname{Buldum}} 
\author{Jian Ping \surname{Lu}\email[Email address]{: jpl@physics.unc.edu}} 
\affiliation{Department of Physics and Astronomy, The University of North 
Carolina at Chapel Hill, Chapel Hill, NC, 27599}
\date{\today}

\begin{abstract}
Quantum interference has dramatic effects on electronic 
transport through nanotube contacts. In optimal configuration the intertube 
conductance can approach that of a perfect nanotube ($4e^2/h$). 
The maximum conductance increases rapidly with the contact length up to 10 nm,
 beyond which it exhibits long wavelength oscillations. 
This is attributed to the resonant cavity-like interference phenomena in the 
contact region. For two concentric nanotubes symmetry breaking can reduce
the 
maximum intertube conductance from $4e^2/h$ to $2e^2/h$. The phenomena 
discussed here can serve as a foundation for building nanotube electronic 
circuits and high speed nanoscale electromechanical devices.
\end{abstract} 

\pacs{73.63.Fg, 73.63.Rt, 73.22.-f, 73.21.Hb}

\maketitle

Carbon nanotubes have electronic and mechanic properties \cite{saito}
that make them excellent candidates for nanoscale electronic
circuits. Simple devices such as diodes \cite{hu99}, single 
electron transistors \cite{tan97,bock97}, field effect transistors 
\cite{tan98,mart98} and 
elementary electronic circuits \cite{bach01} have been built. 
Understanding the electronic transport through nanotube 
contacts will be essential for more complex nanotube circuits. 
Several nanotube/metal contacts 
 \cite{frank98,bach98,anan00,paul00} and 
intra-nanotube junctions 
\cite{chicoprb,men97,tam97,marco99,treb99_2,papa00,sati00,fer00} have 
been studied theoretically and experimentally. The possibility of using 
nanotube/nanotube contacts for electronic devices has been suggested 
\cite{rueck00,bul01,fuhsci,tom00,pos00}.  
In this paper we examine the electronic transport between two  
nanotubes, in parallel and in concentric geometries.
The variation of conductance with the tube chirality and the contact length 
is investigated. Characteristic rapid oscillations in 
conductance, related to the nanotube atomic structure and the Fermi 
wavelength of the conduction band, are found\cite{bul01}. 
In general the inter-tube conductance is small when two tubes have
different chirality. However, for 
{\it armchair/armchair} or {\it zigzag/zigzag} the contact conductance is 
significant and can approach $4e^2/h$ (the conductance of a perfect nanotube)
when the contact length is of the order of 10 nm. Further increase in contact
length reveals long wavelength oscillations. This is attributed to the 
quantum interference of the wavefunction in the contact region, which creates
 resonant cavity-like states. Such oscillations were found in a recent STM 
experiment \cite{liang01}. For two concentric nanotubes it is
found that the maximum intertube conductance is either $4e^2/h$ or
$2e^2/h$ depending on the symmetry of the nanotubes.

The electronic structure of the nanotubes is modeled by a 
simple $\pi$-orbital tight binding hamiltonian\cite{saito}, taking into 
account the nanotube curvature. The quantum conductance
is calculated using 
the Landauer-B\"uttiker formula, with a recursive Green-functions technique 
\cite{marco99,bul01}. This approach has been shown to provide
 good agreement with experimentally measured electronic
structure and transport properties \cite{lieb02,liang01,marco99,bul01}.
Electron-electron interactions play an important role in electronic 
transport through carbon nanotubes \cite{bock99}, however they have little 
effect on conductance when the contacts with the leads are highly transparent
\cite{kong01}. In our model the leads are perfect nanotubes to insure
good contact with the structure being investigated, therefore 
electron-electron interactions are not included in the present 
calculations.

For the case of two nanotubes in parallel contact 
\cite{not2}, the distance between nanotubes is fixed to 3.1\AA \ from 
molecular dynamics simulations. The conductance is not sensitive
 to small changes in the intertube distance around this value.
\begin{figure}
\epsfxsize=3.5in \epsfbox{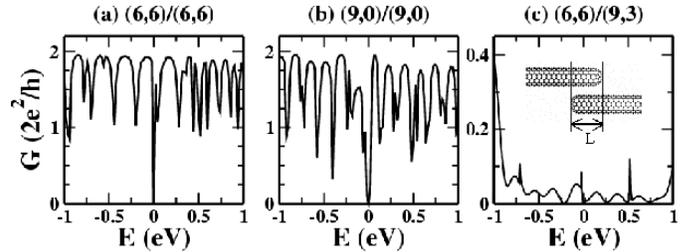}

\caption{Intertube conductance for a (a) (6,6)/(6,6) (b) (9,0)/(9,0) 
(c) (6,6)/(9,3) parallel contact as a 
function of energy for contact lengths (denoted L on the graph) of order of 10
nm. Notice the presence of a small gap at $E=0 eV$ in (a) and (b) due to 
interaction between nanotubes. 
For the same reason the graphs are not symmetric with respect to $E=0 eV$
 point (see also \cite{tnek98}).
}
\end{figure}

\begin{figure*}
\centerline{
\epsfxsize=6.5in \epsfbox{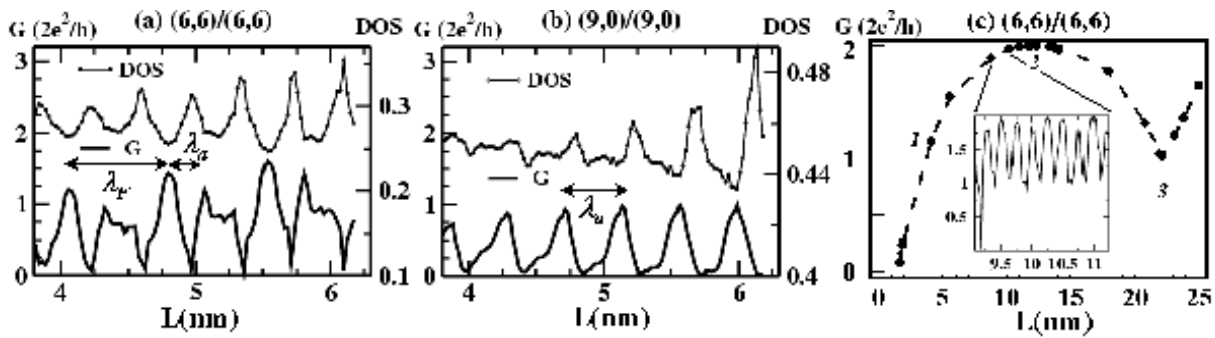}
}
\caption{Conductance and DOS as functions of the contact length at $E=-0.1 eV$
for a (a) (6,6)/(6,6) (b) (9,0)/(9,0) parallel contact. Rapid oscillations with
a period 
of unit cell length are present for both armchair($\lambda_a=2.46$ \AA
($=a_0$)) and zigzag ($\lambda_a=4.26$ \AA($=a_0$)) tubes. Additional 
modulation related to the Fermi wavelength ($\lambda_F=3a_0$) can be seen in 
(a). Due to the fact that $k_F=0$ for zigzag nanotubes such a modulation is 
missing in (b)). Notice the correlation between the peaks in the DOS and the 
minima in the conductance (see the text for discussions).
(c) The upper envelope of conduction oscillation as
a function of the contact length ($L$) at $E=-0.1 eV$, showing long wavelength 
oscillation. The maximum conductance can reach that of a perfect nanotube for 
contact lengths greater than 10nm. Regions labeled by 1, 2, 3 are examined in 
detail in Fig. 2(a), 3(a), 3(c). The solid dots correspond to calculated
local conduction maxima. The insert shows an example of actual dependence
of conductance on contact length.
}
\end{figure*}

Various combinations of metallic nanotubes with different tube size and 
chirality were investigated. The size of nanotubes is found to have no 
substantial effect on the conductance, however it depends sensitively on 
chirality. Shown 
in Fig. 1 are typical examples of intertube conductance as a function of
energy. The best conductance is achieved for
{\it armchair/armchair} or {\it zigzag/zigzag} configurations. For 
{\it armchair/armchair} contacts the conductance can approach that of a 
perfect nanotube, $4e^2/h$ (Fig. 1 (a)). In contrast, it is an order of 
magnitude smaller when the two tubes have different chirality(Fig. 1 (c)). 
This is due to the fact that only {\it zigzag/zigzag} and 
{\it armchair/armchair} contacts allow optimal configurations in which 
delocalized states are present across the contact (see also Fig 3 (a)).
As a function of energy, the conductance exhibits a series of 
minima. Examination of electronic structure revealed that
 these minima correspond to peaks in the density of states (DOS) 
of the contact region. This suggests that resonant backscattering due to the 
quasi bound states formed in the contact area is responsible for the 
conductance dips (see also Fig. 2, 3). Similar effects have been found 
for nanotubes with defects \cite{treb99,choi00,kong01}. 

When examining the dependence of conductance on the contact length we found two
 types of characteristic oscillations [Fig. 2]. At atomic length scale rapid 
oscillation of conductance with the contact length is related to the unit cell
 length ($a_0=2.46$\AA \ for armchair nanotube and $a_0=4.26$\AA \ for zigzag 
nanotube) and is due to the resonant backscattering on the quasibound states 
\cite{choi00} (notice the same correlation between minima in conductance and 
maxima in DOS as in the case of energy dependence). In 
{\it armchair/armchair} contacts additional modulation related to the Fermi 
wavelength is also present \cite{meu99,rub99,ven99}.

As one continuously increases the contact length, the rapid oscillation in 
conductance persists. The envelope of the oscillation shows smooth
variation with the contact length (Fig. 2 (c)). Initially the maximum 
conductance 
increases rapidly with the contact length. The  maximum value approaches 
$4e^2/h$ (value for a perfect tube) for contact lengths of the order of 10nm.
This is surprising considering that the number of quasibound states also 
increases with the contact length. The explanation is that the quasibound 
states are formed mainly for certain local arrangements of the atoms in the 
contact area for which the minima in conductance remain indeed very low 
at any contact length. For other arrangements delocalized 
states are formed which facilitate the conduction. In such cases the DOS has 
values close to those of the perfect tube (see also Fig. 3 (a),(b)). 
Further increase of the contact length shows unexpected long wavelength 
oscillation in conductance (Fig. 2 (c)). Resonant cavity-like interference is 
 responsible for this interesting feature. 
To understand this phenomenon better in Fig. 3 we show 2D contour plots of 
the local density of states (LDOS) as a function of energy and position along 
the contact for fixed contact lengths.
When the conductance is in the vicinity of a global maximum (marked 2 in 
Fig. 2 (c)) the LDOS is small and smooth along the contact (Fig. 3 (a),(b)) 
for most energies.   
In contrast, when the conductance is in the vicinity of a global minimum 
(marked 3 in Fig. 2 (c)), one can observe a clear interference pattern across 
the contact area (Fig. 3 (c)), due to the formation of a resonant cavity
in such a configuration. Such resonance is very similar to that
observed in a recent experiment \cite{lemay01}.

In a semiinfinite tube resonant backscattering of the electrons takes place 
due to the finite end. The incoming wave and the reflected 
wave interfere producing a set of maxima and minima in the LDOS along the 
tube. When two nanotubes are in parallel contact, if the interference maxima 
in the two tubes overlap for a certain energy and contact length, 
the contact exhibits a resonant cavity-like behavior. A standing wave pattern 
in the LDOS along the contact 
can be observed in this case (Fig. 3 (c)) and the conductance has a global 
minimum (area 3 in Fig. 2 (c), Fig. 3(d)). 
When the interference maxima in one tube overlap with the minima in the other 
tube the LDOS in the two tubes are out of phase and the resonant cavity-like 
behavior is destroyed (Fig. 3 (a), (b)). The electrons can be easily 
transmitted through the contact, thus high conductance values 
(area 2 in Fig. 2 (c), Fig. 3(b)).   

\begin{figure}
\epsfxsize=3.5in \epsfbox{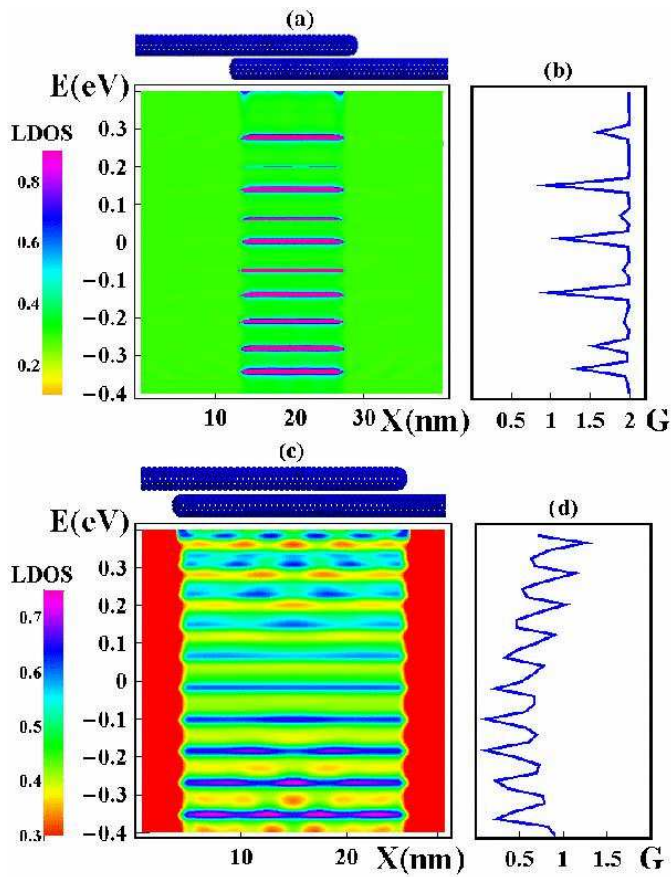} 
\caption{ (a) 2D contour plot of the LDOS (in states/unit cell) along the 
contact as a function of energy and (b) conductance (in units of $G_0=2e^2/h$ 
as a function of energy for a (6,6)/(6,6) contact 
when conductance has a global maximum (L=12nm, area 2 in Fig. 2c). Notice that
 when the conductance is $2G_0$ the LDOS has the same value along the 
nanotubes and across the contact, indicating the continuation of the 
conductance band from one tube to the other. (c) 2D contour plot of the LDOS 
(in states/unit cell) along the contact as a function of energy and (d) 
conductance (in units of $G_0$) as a function of energy for a (6,6)/(6,6) 
contact when conductance has a global minimum (L=22nm, area 3 in Fig. 2c). 
The standing wave pattern for specific values of 
the is very clear. The minima in conductance due to formation of quasibound 
states in the contact area are present in both cases.
}
\end{figure}

As a function of the contact length the long 
wavelength oscillation of the conductance depends on the energy: the further 
the energy is from 0 the smaller is the oscillation wavelength. 
In a crude approximation we can treat this phenomenon as a beat-like 
interference between the incoming wave ($k_1$) and the reflected one ($k_2$) 
when they are in different bands, resulting in a modulation of wavelength 
$ \lambda_i = \frac{2\pi}{k_1-k_2} $.
The dispersion relation for the two conduction bands  of an 
armchair nanotube can be written as $\Delta E=\pm V_0 (1-2\cos \frac{ka}{2})$
\cite{saito}.
 The wave vectors of the electrons located in these bands will be 
$k_{1,2}=\frac{2}{a}(\frac{\pi}{3}\pm \frac{\Delta E}{V_0\sqrt{3}})$ near Fermi
 energy, therefore
$ \lambda_i =\frac{2\pi}{k_1-k_2} = \frac{3\pi V_0d}{2\Delta E}$.
Thus the wavelength of the long range oscillation should be inverse 
proportional to the energy. This can be clearly seen in Fig. 3(c) where as one
moves away from $E=0$ the number of maxima increases with the energy across a
given contact.
Such long range oscillations in LDOS have been observed 
previously in low dimension electron gas systems \cite{yok,jli}. Recent 
experiments on nanotubes also suggest the existence  
of such long wave oscillations \cite{liang01,lemay01,kong01}.

The possibility of using multiwall carbon nanotubes as electric circuits 
components has been suggested since their discovery \cite{dres93,saito93}. 
Recent 
experiments show that building devices out of multiwall nanotubes is now
possible \cite{cum00}, revealing new potential applications \cite{zhe02}. We 
considered a contact built from a multiwall nanotube, consisting of two 
semiinfinite concentric tubes, such that the inner tube can telescope. 
The dependence of the intertube conductance on the 
size and the chirality of the tubes was examined. 
The best conductance is once again achieved by {\it armchair/armchair} and 
{\it zigzag/zigzag} contacts; the other combinations show a conductance 
at least an order of magnitude smaller. Both atomic scale and long wavelength
oscillations are also present. For a given contact length the 
conductance maxima show a steeper initial increase with the contact length due
to the larger contact area of the concentric geometry compared to the parallel
case (Fig. 4). The conductance of the contact depends on the intertube 
distance. The maximum conductance is obtained when the intertube distance 
$d_i$ is around 3.4\AA , the observed interwall distance in multiwall 
nanotubes\cite{ebb92}.

\begin{figure}
\epsfxsize=3.0in \epsfbox{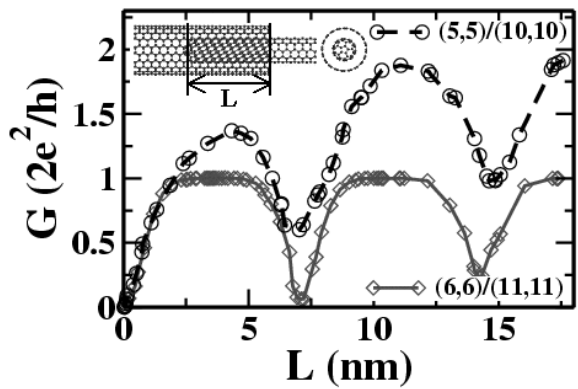}

\caption{Envelope function of the conductance for a (5,5)/(10,10) (dotted 
line) and a (6,6)/(11,11) (full line) concentric geometry contact at
$E=-0.1 eV$. 
}
\end{figure}

Because in the concentric geometry the symmetry axes of the two nanotubes are 
aligned the angular momentum is a good quantum number. This means that an  
electron starting in one of the tubes must 
scatter to a state with the same symmetry in the other nanotube 
\cite{chicoprb} (this is not the case when the two nanotubes are in parallel 
contact, as they do not share a common axis and therefore 
angular momentum is not a good quantum number). Considering the case of 
$(n_1,n_1)/(n_2,n_2)$ nanotube contacts, the two nanotubes have a
$C_{n_1}$ 
and $C_{n_2}$ symmetry respectively. Correspondingly the $\pi$ bands 
have the angular quantum number 0 in both tubes \cite{chicoprb}, while the
 $\pi^*$ bands have $n_1$ and $n_2$ respectively \cite{choi99}. As a 
consequence, the conduction 
channel due to the $\pi$ band remains always open while the 
the conduction channel due to the $\pi^*$ band is open only if the two
nanotubes  have compatible 
rotational symmetry. As an illustration, we show in Fig. 4 the conductance 
for (5,5)/(10,10) (dotted line) and (6,6)/(11,11) (full line) contacts. In the case of 
(5,5)/(10,10) contact the $\pi*$ bands have compatible rotational symmetries 
and the conduction can reach $4e^2/h$. For (6,6)/(11,11) contact the
rotational symmetries are incompatible hence the maximum value reached
by the conductance is $2e^2/h$. This may explain the observation of
only one conductance 
channel in some multiwall nanotubes experiments \cite{frank98}.

In conclusion, nanotube/nanotube contacts exhibit a variety of novel 
interesting phenomena. We found that the best conductance is achieved for 
armchair/armchair and zigzag/zigzag contacts. The conductance maxima increase
 with the contact length and can reach the value for a perfect tube. 
For larger contact lengths a long wavelength oscillatory behavior is found. 
This is attributed to the resonant cavity-like interference phenomena in the 
contact region. For two concentric nanotubes symmetry breaking can reduce
the 
maximum intertube conductance from $4e^2/h$ to $2e^2/h$. The phenomena 
discussed here can serve as a foundation for building nanotube electronic 
circuits and high speed nanoscale electromechanical devices.

\appendix*
\begin{acknowledgments}
This work is supported by the U.S. Army Research Office Grant No.
DAAG55-98-1-0298, the Office of Naval Research Grant No. 
N00014-98-1-0597 and NASA Ames Research Center. We acknowledge computational 
support from the North Carolina Supercomputer Center.
\end{acknowledgments}
\vspace{8mm}

\bibliography{meso.bib}

\begin{thebibliography}{44}
\bibitem[{not()}]{not2}
\bibinfo{note}{Earlier calculations have shown that optimum conductance is
  achieved when two nanotubes are in parallel contact \cite{bul01}.}
\end{thebibliography}

\end{document}